\documentclass[twocolumn]{article}
\usepackage{arxiv}
\usepackage[utf8]{inputenc} 
\usepackage[T1]{fontenc}    
\usepackage{hyperref}       
\usepackage{url}            
\usepackage{booktabs}       
\usepackage{amsfonts}       
\usepackage{amssymb}
\usepackage{amsmath}
\usepackage{nicefrac}       
\usepackage{microtype}      
\usepackage{lipsum}		
\usepackage{graphicx}
\usepackage{natbib}
\usepackage{doi}

\newcommand{\affit}[1]{$^{\mathrm{\textnormal{\textit{#1}}}}$}

\title{Enhancing a high resolution data-driven weather prediction model with surface descriptors} 

\author{
 Åsmund Bakketun \affit{a,}\affit{b,*} \\
  \And
 Håvard Homleid Haugen \affit{a} \\ 
  \And
 Jostein Blyverket \affit{a} \\
  \And
 Thomas Nils Nipen \affit{a} \\
  \And
 Malte Müller \affit{a,}\affit{b} \\
}

\renewcommand{\headeright}{}
\renewcommand{\undertitle}{}

\begin{document}

\twocolumn[
\begin{@twocolumnfalse}
\maketitle
\bigskip

\begin{abstract}
We study the importance of surface characteristics when forecasting near-surface variables with a data-driven weather prediction model. To target the challenge of predicting small-scale weather conditions at high resolution, we introduce a range of surface descriptors in the training of a state-of-the-art data-driven model. The input data includes surface descriptors inherited from the numerical weather prediction model used to produce the training dataset and topographic neighbourhood indices. We found that errors of 2-metre temperature and 10-metre wind speed forecasts were reduced by 1.9\% and 3.0\% respectively compared to a baseline model over the model domain. Over certain surfaces, the improvements were significantly larger. For example, we found a 12\% reduction of temperature mean absolute errors over urban areas when the urban fraction was included in the model input. Furthermore, we investigated how the model responded to removal of glaciers, resulting in an increase of temperature. This indicates that 1) the model produce a physically reasonable response and 2) input datasets can be updated without the need to retrain the model. The latter suggests a great benefit for operational systems as training is expensive compared to running these models. This study highlights the importance of including surface conditions in the prediction of near-surface variables.
\end{abstract}
\bigskip

\vspace{1.5cm}

\end{@twocolumnfalse}
]

\footnotetext[1]{Norwegian Meteorological Institute, Oslo, Norway}
\footnotetext[2]{University of Oslo, Oslo, Norway}

\footnotetext{*Corresponding author: \texttt{asmundb@met.no}}

\section{Introduction}
Data-driven weather prediction models have proven to outperform classical numerical weather prediction (NWP) models in terms of standard verification metrics such as root mean square error and anomaly correlation \citep{keisler_forecasting_2022,lam_graphcast_2023,pathak_fourcastnet_2022, lang2026aifs}. The models show largest improvement at the synoptic scale compared to classical NWP models, while less improvement is seen at the smaller scales \citep{bouallegue_rise_2024, shi_comparison_2025}. Another limitation of the current global data-driven models is the relatively coarse spatial resolution compared to operational NWP models \citep{husain_leveraging_2025}. However, these models are evolving rapidly and resolution will increase in the future \citep{lam_graphcast_2023, husain_leveraging_2025}. 

National weather centres rely on high resolution NWP forecasts to meet the user needs for local weather and hazard warnings. This typically involves running a short range limited area model, using the global forecast as lateral boundaries \citep[eg.][]{muller_characteristics_2017}. Recently, data-driven limited area models or stretched-grid versions have been developed, enabling kilometre-scale weather forecasts for their regions \citep{oskarsson2023graph, nipen_regional_2025,nordhagen_high-resolution_2025, bano-medina_regional_2025}. In addition to providing more accurate weather forecasts, data-driven models significantly reduce the computational cost and time to produce a forecast. With increased resolution, the focus typically shifts from synoptic scale dynamics to near-surface variables like 2-metre temperature, 10-metre wind speed and precipitation amount. 

The surface plays an important role in modelling of the earth system and it is particularly important for capturing the dynamics of the aforementioned variables. For example, surface moisture modulates the portion of energy being released as sensible or latent heat which directly impact the air temperature \citep{seneviratne_investigating_2010}. The amount of energy absorbed by the surface also depend on the albedo of the surface which is varying between different surface types \citep{noilhan_simple_1989}. Wind speed is affected by the surface roughness, which again impacts the heat fluxes \citep{louis_parametric_1979}. Precipitation, wind speed and temperature are all strongly forced by topography \citep{smith_influence_1979}. In NWP systems, complex surface models are used to simulate the lower boundary to account for these processes. The surface schemes rely on extensive datasets of parameters describing the surface types, vegetation and elevation data \citep[e.g.][]{balsamo_revised_2009, monteiro_improvements_2024}. From here on, we refer to such parameters as surface descriptors.

Data-driven models make predictions from the relationships between input and output learned during training. Varied and high quality training data is thus essential for an accurate prediction model. Over surfaces that are covering small fractions of the domain, they might not learn important processes due to under sampling. For example, urban areas, where most people live, cover only a small fraction of the Earth. Urban areas are subject to important processes like the urban heat island effect \citep{oke_energetic_1982}, which is critical to capture in case of heat waves. For the data-driven models to be able to predict local phenomena, location specific information is crucial. There are several approaches to solve this problem. One solution is to add a layer of location-specific, trainable parameters (learnable features) in the model. This method allows the model to consider any local effect without having to explain what causes it. While learnable features can capture the local effects, these parameters tie the model to a fixed grid and prevent transfer to other domains or resolutions. Another option is to provide surface descriptors as input. This allows for flexibility of changing domain, and the potential to adjust and update the surface descriptor maps in the future. Within the current data-driven models, a limited set of surface descriptors are used as input, sometimes only elevation and land-sea mask \citep[e.g.][]{nordhagen_high-resolution_2025}. This raises the question whether this is sufficient for providing accurate forecasts of near-surface variables on high resolution, particularly over under-sampled surfaces.

In this study we assess to what extent additional surface descriptors improve a high-resolution data-driven weather forecasting model's ability to predict small-scale features and local atmospheric phenomena. We extend the set of input variables to the data-driven model and evaluate the impact on near-surface atmospheric variables compared to a reference experiment with the standard input of variables \citep{nordhagen_high-resolution_2025}.

This article is structured into a methods and data section (Section~\ref{seq:method}), results (Section~\ref{seq:results}) followed by a discussion (Section~\ref{seq:discussion}). Conclusions are drawn in Section~\ref{seq:conclusion}.

\section{Methods and Data}\label{seq:method}
In this study, we use the data-driven weather model Bris \citep{nordhagen_high-resolution_2025}. This is a global model with a stretched grid, allowing higher resolution over a focus region and lower resolution elsewhere. At MET Norway, Bris has 2.5km resolution for the Nordic region and 31km elsewhere.

Although the model has a flexible grid that can be changed, the model expects a fixed set of input parameters. Changing this requires a complete retraining of the model, which for Bris amounts to 24,000 GPU-hours \citep{nordhagen_high-resolution_2025} on NVIDIA A100 GPUs. This makes studying the effect of adding different surface parameters prohibitively expensive. To reduce this cost, we keep the existing trained Bris model and train a separate decoder for 2-metre temperature and 10-metre wind speed, reducing the cost to around 2,500 GPU-hours.

In this section, we describe the Bris architecture and how the second decoder works (Section~\ref{sec:model}), the surface descriptors that we investigate as inputs to the second decoder (Section~\ref{sec:forcings}), and the configuration of variables used in the experiments we run (Section~\ref{sec:configurations}).

\subsection{Model}\label{sec:model}
\begin{figure}
    \centering
    \includegraphics[width=0.9\linewidth]{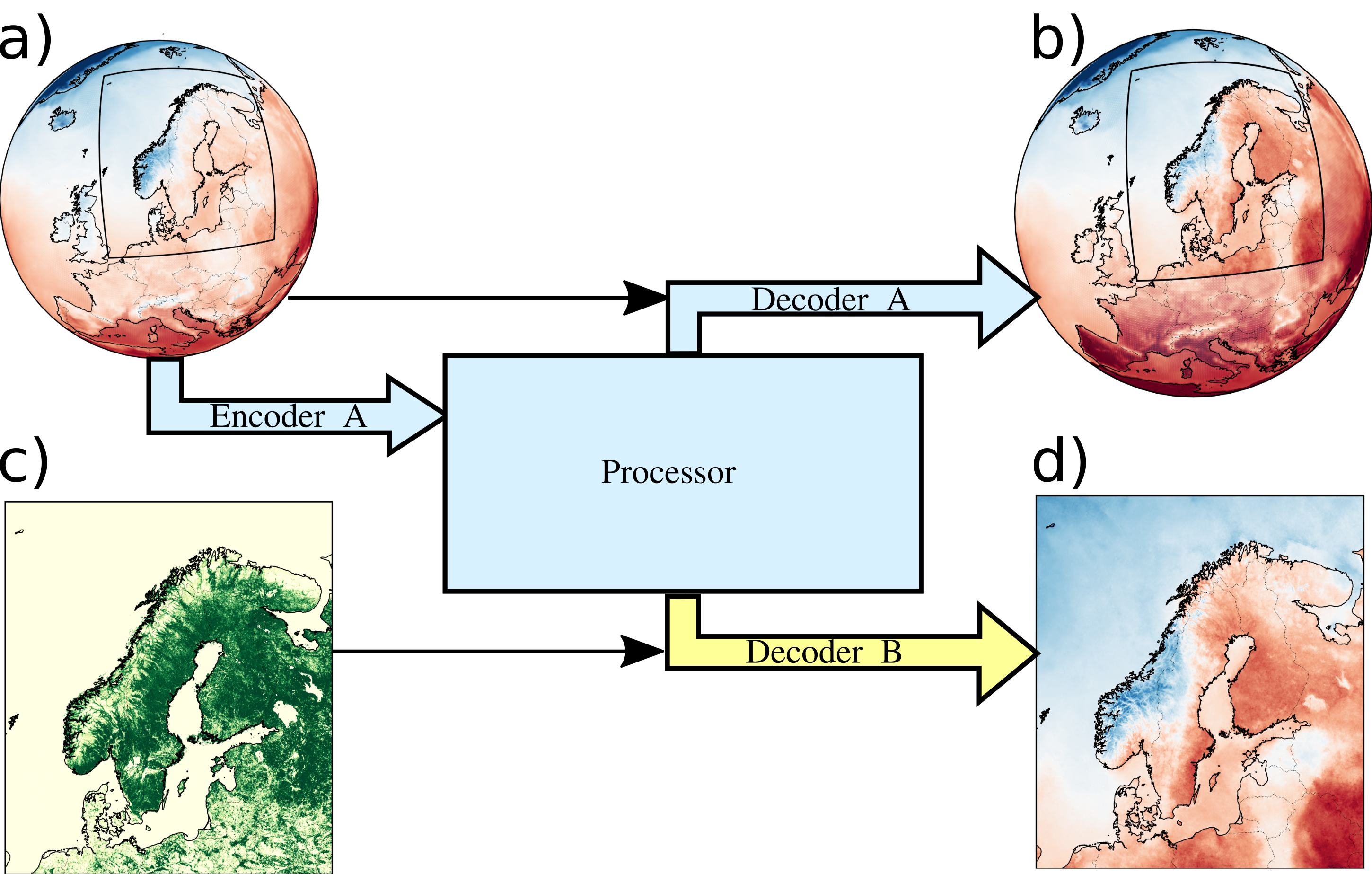}
    \caption{Model component diagram with examples of input and output data. Input and output in the base model referred to as Bris in the text is shown in a) and b) represented by 2-metre temperature fields on the stretched grid with high resolution over the Nordics. c) shows fraction of forest representing surface descriptor input to the secondary decoder which only covers the Nordics. d) shows 2-metre temperature field representing output from the secondary decoder covering the Nordics. The base model is marked in blue, while the secondary decoder (Decoder B) is marked in yellow. During training, only weights associated with the yellow part of the model are optimized, the other components are kept constant from the pre-trained model. The thin arrows indicate the direct connection between model input and decoder.}
    \label{fig:input_output}
\end{figure}

Bris is a global probabilistic weather prediction model with increased spatial resolution over a limited area of interest \citep{nordhagen_high-resolution_2025}. Using a stretched grid, the model is trained on a combination of global data from the ERA5 reanalysis \citep{hersbach_era5_2020} and IFS operational analysis, both at 0.25 degree horizontal resolution, and operational analysis from the MetCoOp ensemble prediction system (MEPS) at 2.5km resolution. This is illustrated in Fig.~\ref{fig:input_output}a-b). The model is trained in multiple stages, where it is first pre-trained on 40 years of global data followed by a shorter fine tuning on 3 years of combined high resolution and global data.

Model input and output can be categorized into three groups. \textit{Prognostic} variables are those that are both inputs and outputs of the model. The model learns how to predict a change in these variables from one time step to the next. Variables that are only produced as outputs are called \textit{diagnostics}. For example, precipitation is a diagnostic variable in Bris. Variables that are only input to the model are called \textit{forcings}. Forcings provide information to the model that are well known or easily obtained, for example local time, or geographical location. Forcings can be separated into static and dynamic. Dynamic forcings include local time, Julian day and insolation which can be calculated on the fly. Static forcings do not change during a forecast and can be provided from a dataset. Static forcings include topography, land-sea mask, latitude and longitude. 

Bris is a graph neural network model consisting of an encoder, a processor and a decoder. The encoder maps the input data from physical variable space and resolution to a latent representation on a processor mesh with lower spatial resolution but a larger feature (variable) dimension. The processor consists of a number of message passing steps, where information is passed between nodes on the processor mesh. Finally, the decoder maps the latent representation of the data back to physical variable space and resolution. This process generates a 6-hour forecast and it can be repeated auto-regressively to generate a forecast of arbitrary length. The encoder, processor and decoder all send information between two sets of nodes which are connected by a set of edges. A graph transformer with multi-head attention is used to aggregate information from source nodes in each of these message passing steps \citep{shi2020masked}. For a detailed description of the Bris architecture and hyper-parameters, see \citet{nordhagen_high-resolution_2025}.

In this work we extend the Bris model with an additional decoder, which can output to an arbitrary set of grid points and parameters independent of the base model. The setup is sketched in Fig.~\ref{fig:input_output}, and in our experiments the added decoder outputs 2-metre temperature and 10-metre wind speed over the Nordic high resolution area of the stretched grid. During training we load the weights from a pre-trained single decoder Bris model. The weights for the additional decoder are randomly initialized and fine-tuned while the pre-trained weights of the base model are kept constant. Training parameters are shown in Table~\ref{tab:training_schedule}. This workflow allows us to train several different decoders for different sets of parameters without retraining the full model from scratch. In this work we investigate the impact of adding additional forcing parameters. In the regular single-decoder Bris model, these fields are passed to the decoder through two routes; (i) they are propagated through the processor and (ii) they are embedded in the query feature vector of the decoder. In the additional decoder, only the latter connection is used for the additional forcings. Because these forcings contain local high resolution information about the output grid, it is assumed that they have the most impact through the second connection. 

\begin{table*}[h]
\centering
\caption{Training schedule and optimization hyper parameters.\label{tab:training_schedule}}
\begin{tabular}{l l}
\hline
\textbf{Parameter} & \textbf{Value / Description} \\
\hline
Optimizer &  AdamW \\
Learning rate & $1\times10^{-3}$ (initial) \\
Learning rate schedule & Warmup + cosine decay \\
Warmup steps & 100 \\
Batch size & 16 \\
Number of epochs/steps & 70/15000 \\
Loss function & CRPS (Eq.~\ref{eq:loss}) \\
Hardware & 64 A100 GPUs \\
Computational cost & 2560 GPU hours\\
Training period & 1 January 2020 - 1 June 2022 \\
Validation period & 1 June 2022 - 1 June 2023 \\
\hline
\end{tabular}
\end{table*}

The model is trained to be probabilistic, where stochasticity is introduced by injecting Gaussian noise into the latent space representation of the model in the processor. For each training sample an ensemble of \(M\) forecasts \(\{x_i\}_{i=1}^M\) is generated by multiple copies of the model with different random noise patterns. The forecast is then scored using the \emph{almost fair} Continuous Ranked Probability Score (CRPS) as the loss function, which for a given variable \(v\) and output point \(p\) is given by

\begin{equation}
\begin{split}
    \mathcal{L}(\{x_{v,p}\}_{i=1}^M) = & \frac{1}{M}\sum_{i=1}^M |x_i - y| \\
    & - \frac{1-\epsilon}{2M(M-1)}\sum_{i=1}^M \sum_{j=1}^M |x_i - x_j|,
    \label{eq:CRPS_point}
\end{split}
\end{equation}
where \(y\) is the target. The almost fair CRPS is a linear combination of regular and fair CRPS, designed to avoid a numerical instability in the fair CRPS with small ensemble sizes \citep{lang2026aifs}. The balance between the two terms is decided by \(\epsilon\), where \(\epsilon=0\) gives fair CRPS and \(\epsilon=1/M\) gives regular CRPS. In this work we use \(M=2\) and \(\epsilon = 0.025\), consistent with recent literature \citep{lang2026aifs, nordhagen_high-resolution_2025}. 

Equation~\eqref{eq:CRPS_point} is the loss per variable and output point, so the total loss of a prediction is given by
\begin{equation}
    \mathcal{L} = \sum_{v, p} w_{v,p} \mathcal{L}(\{x_{v,p}\}_{i=1}^M),
    \label{eq:loss}
\end{equation}
where the weighting factor \(w_{v,p}\) is a hyper-parameter. The loss is only computed for points and variables in the additional decoder, since the weights of the base model are frozen. We assign equal weight to all output points because the output grid of the decoder has uniform spacing. The variable weighting used is the same as that of the base model: 1.0 for temperature and 0.1 for zonal and meridional winds. 

In this work we use the Anemoi framework, an open source data-driven weather prediction system developed by the ECMWF and member state organizations \citep[software]{anemoi_contributors_ecmwfanemoi-core_2026}. Anemoi consists of building blocks which provides the full chain from training to prediction.

\subsection{Forcings}\label{sec:forcings}
The MEPS dataset is based on operational NWP simulations using the HARMONIE-AROME model configuration \citep{bengtsson_harmoniearome_2017, frogner_harmonepsharmonie_2019}. This system use the externalised surface model (SURFEX) on the lower boundary \citep{masson_surfexv72_2013}. SURFEX represent a grid cell through separate energy and mass budgets for four tiles: sea, inland water, town and nature. The nature tile, simulated by the ISBA scheme \citep{noilhan_isba_1996, boone_interactions_2017}, is further separated into patches. Each patch in the nature tile represent different surface and vegetation classes. The number of patches is configurable in up to twelve different vegetation classes. In the MEPS configuration used for training Bris, two patches are used, representing high and low vegetation. In the coupling with the atmospheric component, the weighted average fluxes from each tile is communicated, where the contribution from the nature tile is again a weighted average of fluxes from each patch. SURFEX uses a physiography dataset that includes information about soil type, vegetation type, root depth, leaf area index, spectral albedo, fraction of high and low vegetation, subgrid topography, fraction of town, lake, sea and land and more \citep{masson_global_2003, faroux_ecoclimap-iieurope_2013}. 


Since the target model is given this information, we let the first set of additional forcings be based on the physiography dataset. The surface descriptors are presented in Table~\ref{tab:T1} and is a subset of the SURFEX model input. 
In NWP models, each grid cell communicates throughout the model integration. In Bris, message passing is performed in the processor at relatively coarse spatial resolution, but the decoder does not directly connect neighbouring grid cells. We thus introduce topographic descriptors that give neighbourhood context in each grid point listed in Table~\ref{tab:T2}. A kernel diameter of 12.5km is used, which gives 5 grid point wide neighbourhoods. The topography descriptors are derived from the surface elevation field using the filtering software from \citet[Topo-Descriptors][software]{noauthor_meteoswisstopo-descriptors_2026}.

\begin{table*}[h!]
\caption{Table of baseline forcings}\label{tab:T0}
\centering
\begin{tabular}{ll}
\hline
\textbf{Name} & \textbf{Comment} \\
\hline
lsm & land-sea mask\\
ZS (z) & surface elevation (m)\\
cos Julian day & cosine of Julian day \\
cos latitude & cosine of latitude \\
cos local time & cosine of local solar time \\
cos longitude & cosine of longitude  \\
insolation & incoming solar radiation \\
sin Julian day & sine of Julian day \\
sin latitude & sine of latitude  \\
sin local time & sine of local solar time  \\
sin longitude & sine of longitude  \\
\hline
\end{tabular}
\end{table*}

\begin{table*}[h!]
\caption{Table of surface descriptors from SURFEX}\label{tab:T1}
\centering
\begin{tabular}{ll}
\hline
\textbf{Name} & \textbf{Comment} \\
\hline
Forest & fraction of forest cover (patch 2) \\
Tree height & mean tree height (m)  \\
Avg. ZS & average surface elevation (m)  \\
Clay & clay fraction of soil  \\
Sand & sand fraction of soil \\
Glacier & fraction of glacier \\
Nature & fraction of natural land  \\
NPI & natural land fraction position index \\
Sea & fraction of sea \\
Town & fraction of urban/town  \\
Water & fraction of inland water \\
Max ZS & maximum subgrid elevation (m) \\
Min ZS & minimum subgrid elevation (m) \\
Sil ZS & subgrid silhouette \\
SSO Anis & subgrid orography anisotropy \\
subgrid slope x & slope in x-direction (m/m) \\
subgrid slope y & slope in y-direction (m/m) \\
\hline
\end{tabular}
\end{table*}

\begin{table*}[h!]
\caption{Table of topography descriptors}\label{tab:T2}
\centering
\begin{tabular}{ll}
\hline
\textbf{Name} & \textbf{Comment}\\
\hline
ZS SN Derivative
  & north–south elevation derivative \\

ZS Std
  & standard deviation of elevation (m) \\

ZS SX 180 
  & horizon angle in 180° direction (south-facing) \\

TPI
  & topographic position index \\

Valley x 
  & valley orientation vector (x component) \\

Valley y 
  & valley orientation vector (y component) \\

ZS WE Derivative 
  & west–east elevation derivative  \\
\hline
\end{tabular}
\end{table*}

\subsection{Decoder configurations}\label{sec:configurations}
In this study we train four separate decoders. 1) a baseline (BASELINE) decoder with the same forcings as the base model Bris. 2) SFX, which uses the forcings in Table~\ref{tab:T1}. 3) TOPO, using the forcings in Table~\ref{tab:T2}. 4) COMBINED, with forcings from both tables. All decoders include the forcings in BASELINE (Table~\ref{tab:T0}). The decoders output predictions onto the MEPS grid. First we train BASELINE so that the validation metric is similar to that of the base model. When a satisfactory training schedule is found, the other decoders are trained following the same schedule, that is, with the same data, number of iterations and learning rate given in Table~\ref{tab:training_schedule}. For training data-driven weather prediction models, a common strategy is to train on longer lead times (several forecast steps) known as rollout or multi-step training \citep[e.g.][]{nordhagen_high-resolution_2025, cho_deep_2026}. While rollout training can improve model forecasts at longer lead times, our study focuses on evaluating the impact of surface descriptors. The primary objective of our study is to assess the model's response rather than optimizing performance at extended lead times, and thus we opted to not use rollout training in this study.

To evaluate overall characteristics of the different decoders, we run one year inference (1 June 2022 to 1 June 2023) with 4 initialization times daily (00, 06, 12, 18 UTC). Since we compute averages over a year, we assume that stochastic noise cancels out, and we thus run a single member ensemble to limit the computational cost. For the reasons above, we chose to output one time step (6 hours) for the simulated year. For more detailed assessments, we run longer lead time inference for the relevant periods. As ground truth, we use the validation target dataset referred to as MEPS from here. 

\section{Results}
\label{seq:results}
\subsection{BASELINE characteristics}
\begin{figure}
    \centering
    \includegraphics[width=\linewidth]{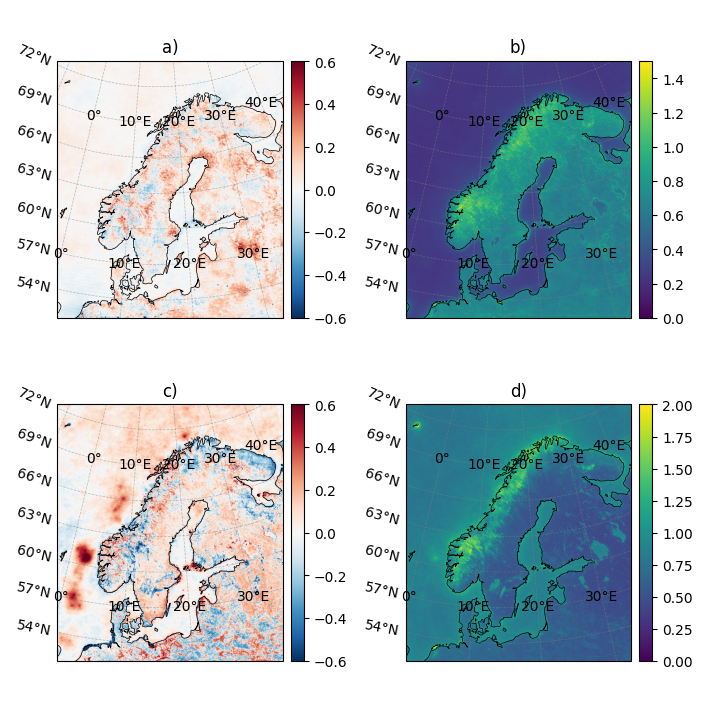}
    \caption{2-metre temperature mean difference (a) and mean absolute difference (b) and 10-metre wind speed mean difference (c) and mean absolute difference (d) between BASELINE and MEPS (target) for the test year (June 2022 - June 2023).}
    \label{fig:baseline_mean_maps}
\end{figure}

Pointwise mean errors (bias) and mean absolute errors (MAE) between the BASELINE 6-hour forecast and MEPS analysis is plotted for 2-metre temperature and 10-metre wind speed in Fig.~\ref{fig:baseline_mean_maps}. For 2-metre temperature, the BASELINE decoder has large scale patterns of warm and cold biases. While these patterns are overall large and smooth, sharper structures are seen within. In terms of MAE, largest errors are seen over mountain regions in Norway, along coastlines including lakes. Over sea, the bias and MAE are smaller compared to those over land.

For 10-metre wind speed we first note large blobs of positive values in the north sea and along the Norwegian coast. The high values are collocated with offshore oil rigs equipped with wind gauges assimilated in MEPS. Over land the patterns have sharper gradients compared to those of 2-metre temperature. We note the negative bias in the northern part of the Kola peninsula extending into the Finnmark plateau. Large areas of negative values are also seen over the southern part of the domain. Sharp high positive values are seen inside the same southern area. Large parts of Sweden and the eastern inland areas have positive values. For MAE, the mountain regions in Norway stands out with largest errors. The errors over sea are higher compared to over inland areas with highest values closer to the coast. 

\begin{figure}
    \centering
    \includegraphics[width=\linewidth]{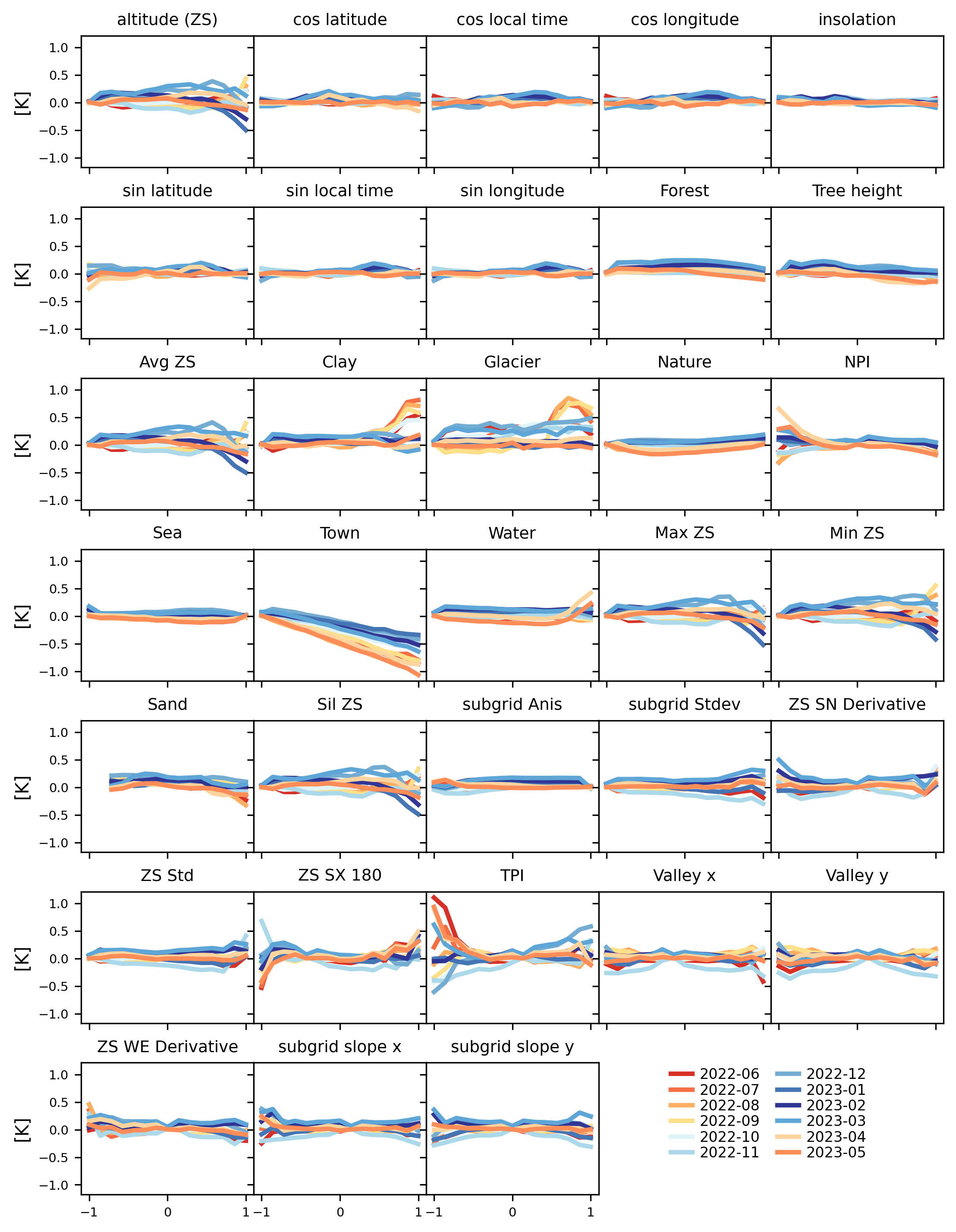}
    \caption{Mean 2-metre temperature differences between BASELINE and MEPS (target) are shown as a function of normalized surface descriptor values. The surface descriptor range is divided into discrete bins, and the mean error for each bin is computed from all grid points whose descriptor values fall within that interval. Monthly mean errors are shown as separate lines, illustrating the seasonal variation in the error–descriptor relationship.}
    \label{fig:bias_vs_forcing_matrix_2t}
\end{figure}

To investigate how 2-metre temperature and 10-metre wind speed varies over different surfaces we show the mean errors of BASELINE as a function of different surface descriptors. The mean errors for 2-metre temperature are shown in Fig.~\ref{fig:bias_vs_forcing_matrix_2t}. For many surface descriptors the lines are flat, indicating no or weak systematic relationship between the forcing value and 2-metre temperature bias. We note the the increase in errors for higher elevation, seen for several topography descriptors (altitude, Max ZS, Min ZS, Avg ZS, Sil ZS). The most clear relationship is seen for Town indicating a consistent cold bias over urban areas. Warm bias is seen for Glacier, particularly during some summer months. During summer, positive bias is also seen for high fractions of Clay. A spread in monthly 2-metre temperature bias is seen for low values of the NPI indicating a sea, lake or urban area being close to natural land pixels. For the topography descriptors, there is a tendency of larger errors for larger magnitudes of the descriptor, particularly ZS SN Derivative, ZS SX 180 and TPI. 

\begin{figure}
    \centering
    \includegraphics[width=\linewidth]{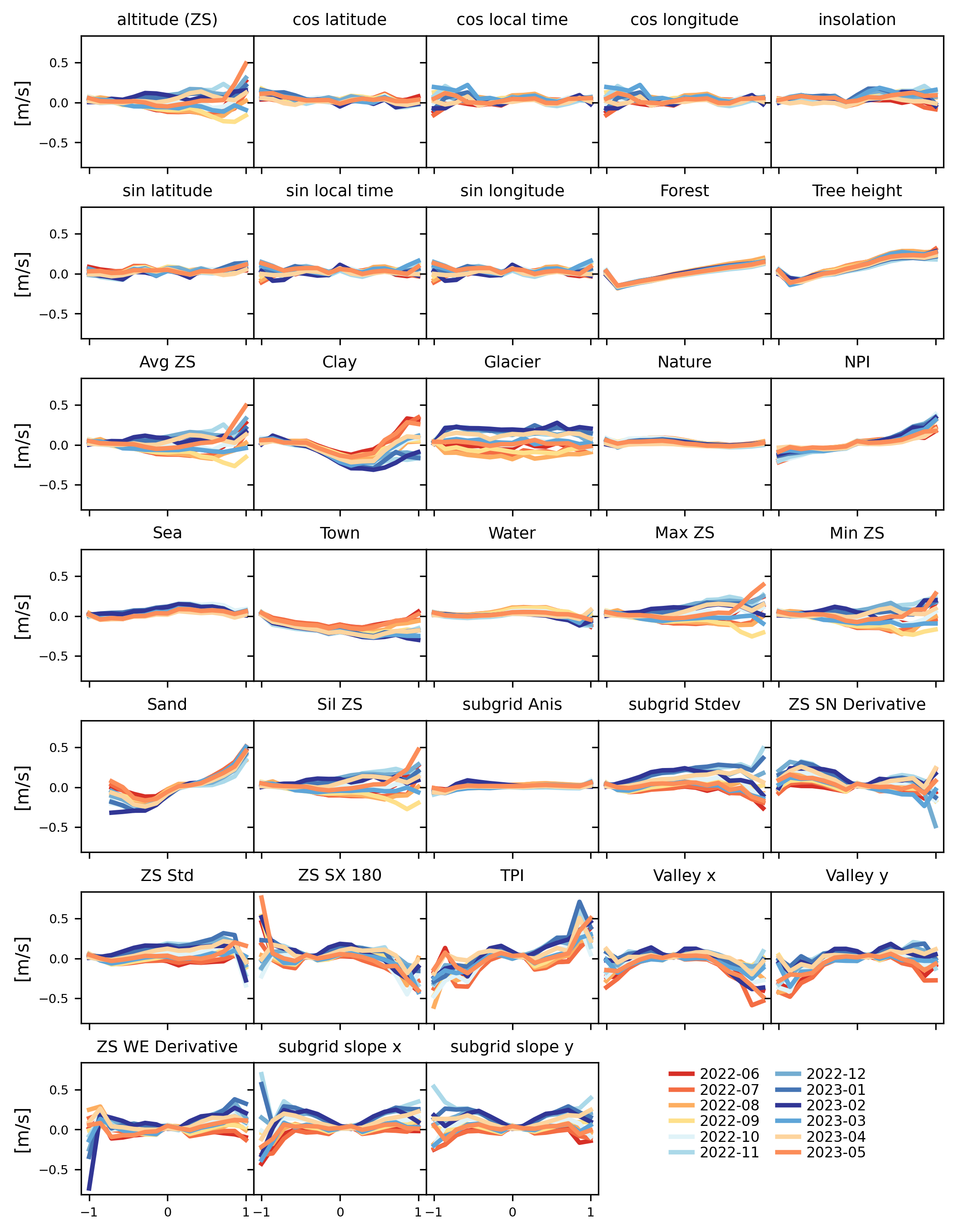}
    \caption{Same as Fig-~\ref{fig:bias_vs_forcing_matrix_2t} but for 10-metre wind speed}
    \label{fig:bias_vs_forcing_matrix_ws}
\end{figure}

Similar relationships are shown for 10-metre wind speed in Fig.~\ref{fig:bias_vs_forcing_matrix_ws}. 10-metre wind speed shows similar patterns as 2-metre temperature, especially larger errors for higher elevation. Forest fraction and Tree height have a consistent relationship with negative 10-metre wind speed bias for low Forest and low Tree height, and increasing bias with higher Forest fraction and Tree height. NPI also has a consistent relationship with negative bias for negative NPI and positive bias for positive NPI. This indicates that 10-metre wind speed is too low over sea, lake or urban areas which are close to natural land points, and that wind speeds are too high over land close to sea, lake or town. The curvature of the lines for Sea, Lake and Town indicates the same relationship as the errors are increased at the interface between land and water. Over Glacier, there is a seasonal dependency on the sign of the mean errors with lower bias during spring and autumn and higher during winter. Compared to 2-metre temperature, there are larger errors relative to the topography descriptors. For TPI, there is a positive trend from negative to positive TPI values. This suggests that too low 10-metre wind speeds are predicted at valley pixels and too high 10-metre wind speed is found at peaks and ridges. Subgrid slopes and topography derivatives shares the over all patterns with low bias in steep terrain during summer and high bias during winter months. 

\subsection{Summary scores}
\begin{figure}
    \centering
    \includegraphics[width=\linewidth]{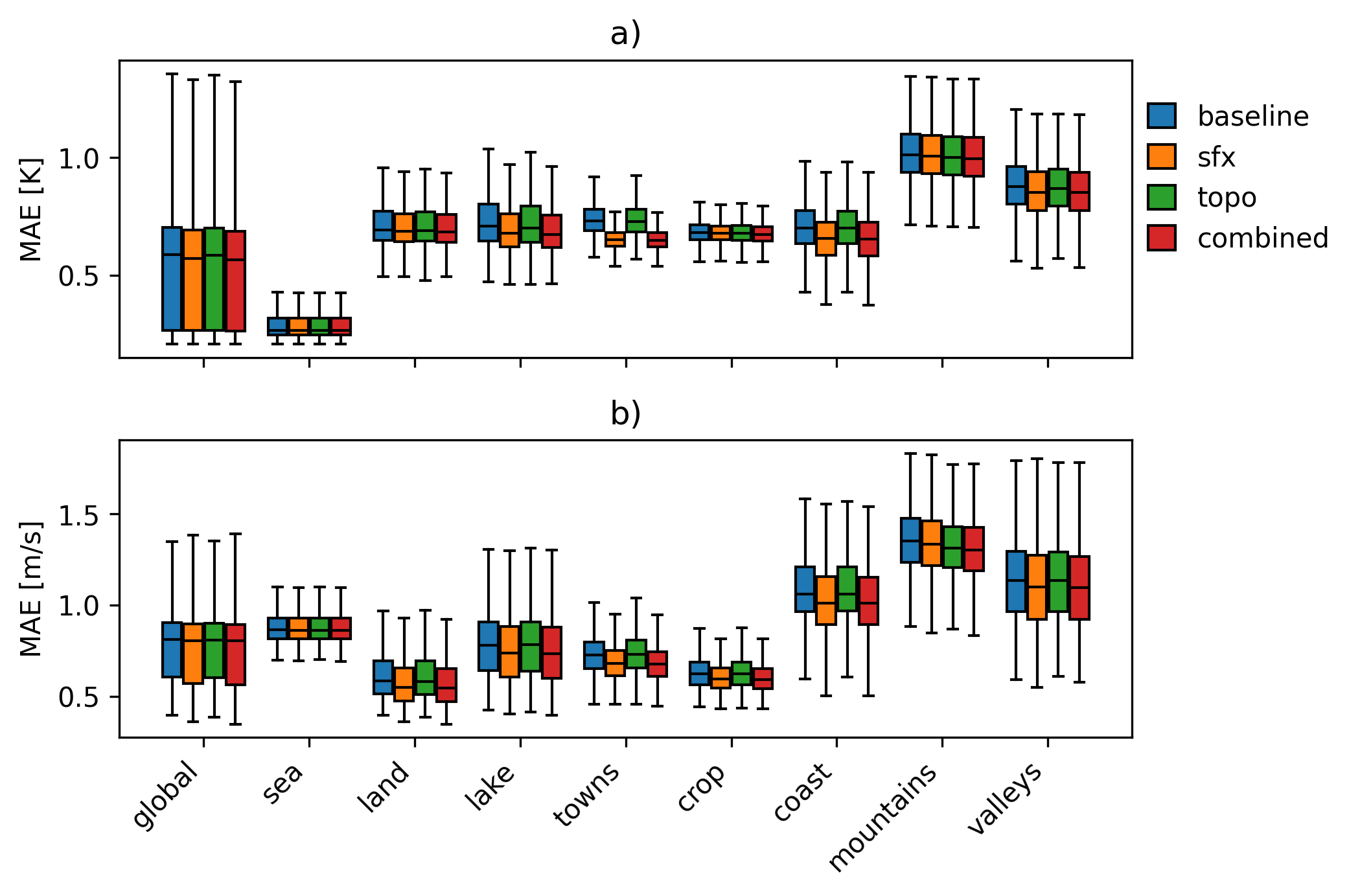}
    \caption{MAE against MEPS for (a) 2-metre temperature and (b) 10-metre wind speed over a selection of surface types. The distribution of MAE is depicted using box and whiskers for each decoder across different surface types. The boxes represents the 25-75\% percentiles (interquartile range, IQR) and whiskers indicates the $1.5\times \mathrm{IQR}$ extensions}
    \label{fig:mae_by_mask}
\end{figure}

To evaluate the overall impact on performance when including surface descriptors we compute MAE over the domain and a selection of potential high impact areas presented in Fig.~\ref{fig:mae_by_mask}. Across the domain, errors are reduced by 1.9\% and 3.0\% for 2-metre temperature and 10-metre wind speed in COMBINED compared to BASELINE. Over sea, we recognize the low 2-metre temperature errors and higher 10-metre wind speed errors compared to over land (Fig.~\ref{fig:baseline_mean_maps}) for all decoders. For other surfaces the pattern is similar: SFX and COMBINED have smaller errors than BASELINE and TOPO for both 2-metre temperature and 10-metre wind speed. The largest difference between the decoders is seen over town where COMBINED has a reduction in MAE of 11.8\% for 2-metre temperature and 7\% for 10-metre wind speed. For coastal areas, COMBINED has a reduction of 6.5\% for 2-metre temperature and 4.8\% for 10-metre wind speed errors. Over land COMBINED has a 5.6\% reduction in 10-metre wind speed errors. The pattern over mountains is somewhat different, where TOPO has smaller errors (2.5\% reduction) compared to SFX (1.2\% reduction) for 10-metre wind speed. 

\subsection{2-metre temperature over towns}

\begin{figure}
    \centering
    \includegraphics[width=\linewidth]{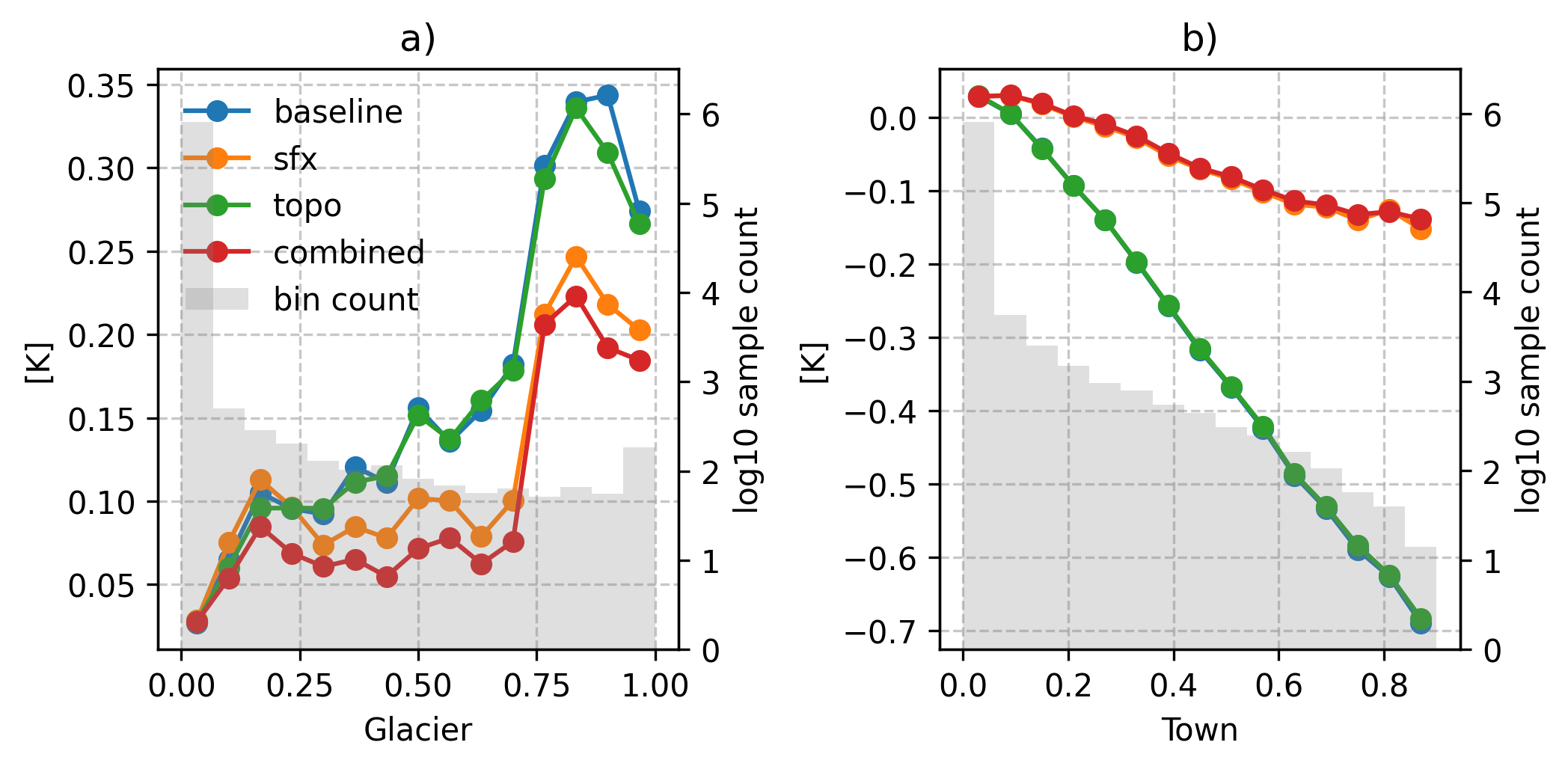}
    \caption{Mean errors (bias) of 2-metre temperature plotted against the fraction of Glacier (a) and the fraction of Town (b) for the four decoders. Grey bars indicate the number of grid points in each bin. Dots represent the bias calculated over grid points with surface descriptor values within the corresponding bin.}
    \label{fig:binned_temperature_detail}
\end{figure}

\begin{figure}
    \centering
    \includegraphics[width=\linewidth]{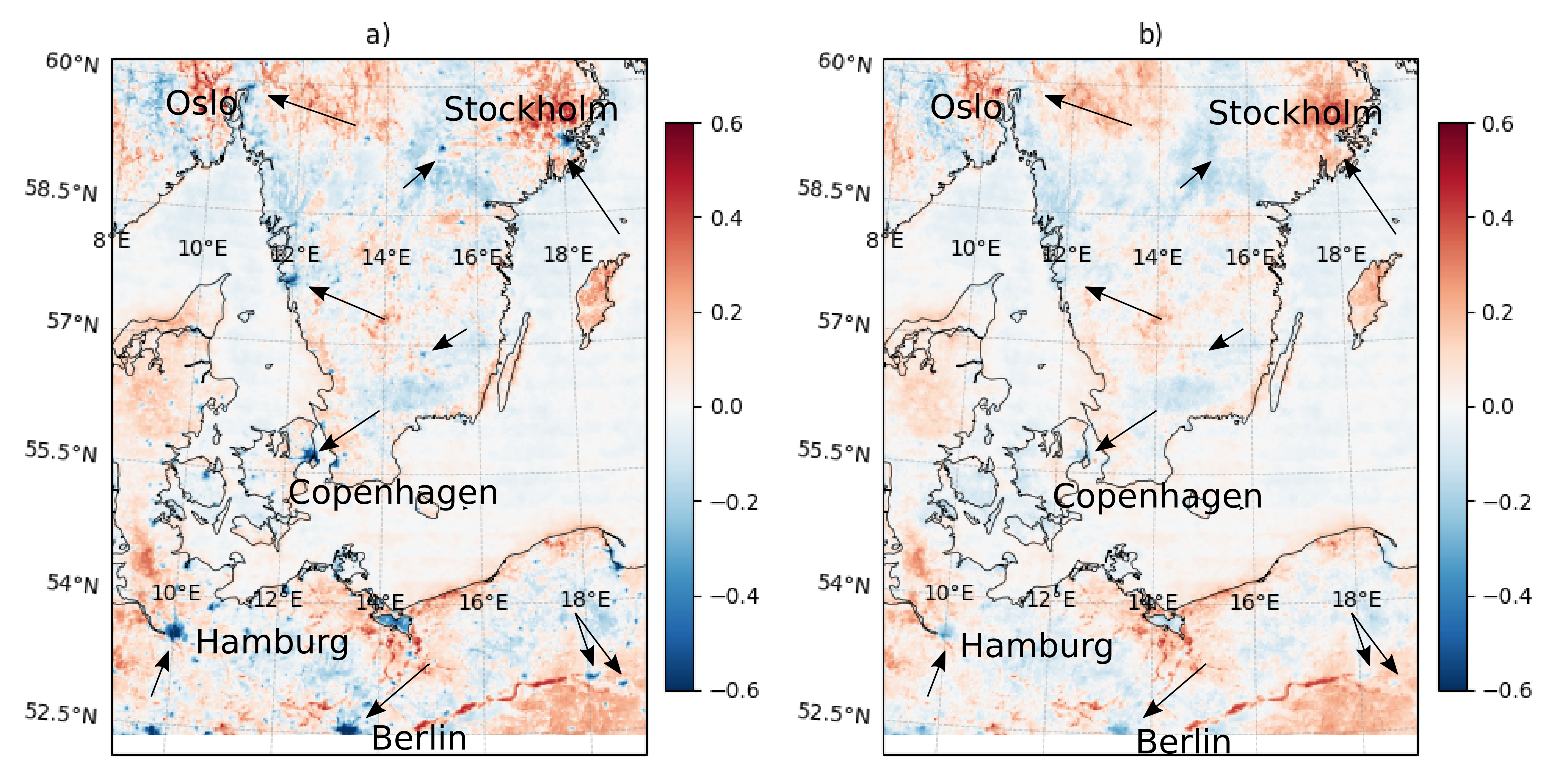}
    \caption{2-metre temperature bias (zoomed version of  Fig.~\ref{fig:baseline_mean_maps}a) for BASLINE (a) and SFX (b). Arrows indicates a non-exhaustive selection of cities with a cold bias in BASELINE.}
    \label{fig:2t_bias_zoom}
\end{figure}

Figure~\ref{fig:bias_vs_forcing_matrix_2t} indicates a consistent relationship between 2-metre temperature bias and town fraction. The effect of including the town fraction forcing in the decoder is shown in Fig.~\ref{fig:binned_temperature_detail}b). Here the town aware decoders (SFX and COMBINED) have reduced the cold bias magnitude to a fifth compared to BASELINE (and TOPO). Figure~\ref{fig:2t_bias_zoom} shows the 2-metre temperature bias on a map over the southern part of the domain where most of the large cities are located. The map indicates that the negative bias is reduced overall, but the largest cities still have a slight cold bias in SFX. 


\subsection{10-metre wind speed over high and low vegetation}
\begin{figure}
    \centering
    \includegraphics[width=\linewidth]{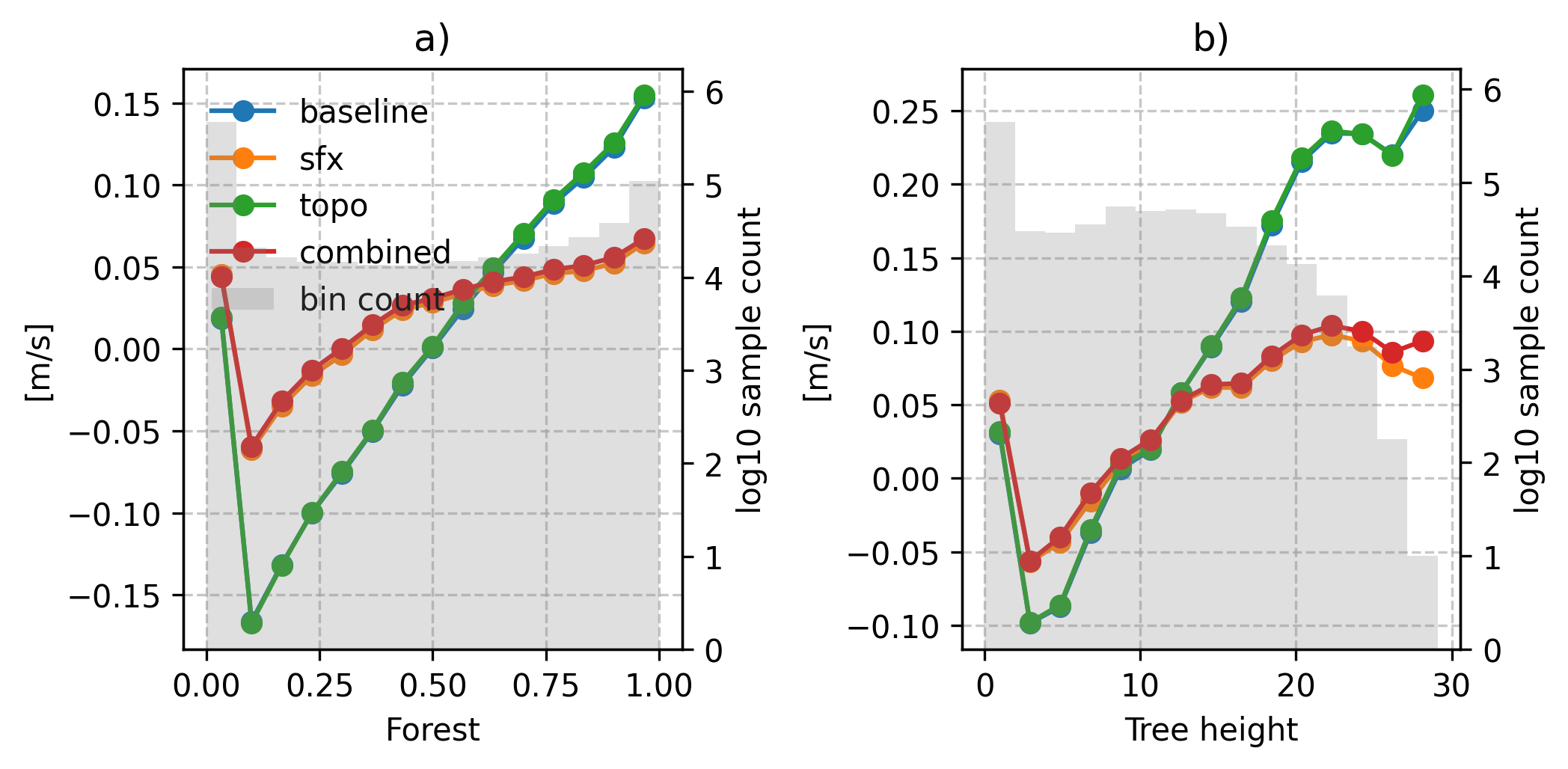}
    \caption{Similar to Fig.~\ref{fig:binned_temperature_detail} but for 10-metre wind speed bias plotted against the fraction of Forest (a) and Tree height (b).}
    \label{fig:binned_bias_ws_forest}
\end{figure}

In Fig.~\ref{fig:bias_vs_forcing_matrix_ws} we find that 10-metre wind speed is too high over forested areas and too low over open land in BASELINE. Figure~\ref{fig:binned_bias_ws_forest} indicates that providing forest fraction and tree height forcings to the decoders (SFX and COMBINED), reduces both the negative and positive bias for low and high vegetation respectively. 

\subsection{2-metre temperature over glaciers}
Figure~\ref{fig:binned_temperature_detail}a) shows reduced 2-metre temperature errors for the SFX and COMBINED decoders for grid points with higher glacier fractions. COMBINED has slightly smaller errors than SFX which suggest a benefit from including the topo descriptors. However, the same benefit is not seen when comparing TOPO and BASELINE. 

\begin{figure}
    \centering
    \includegraphics[width=\linewidth]{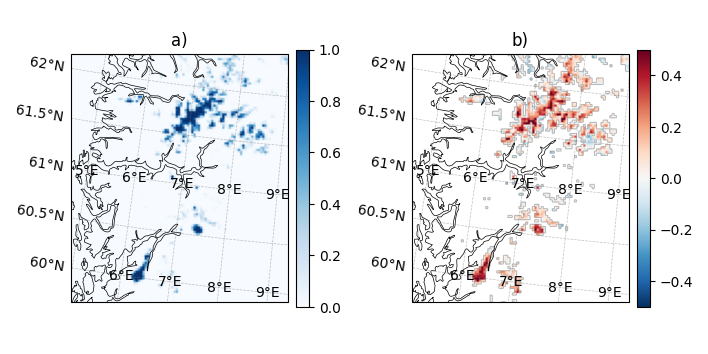}
    \caption{Map of original Glacier field (a) and 2-metre temperature mean difference between SFXG0 and SFX during August 2022 (b). The map is zoomed in over western Norway}
    \label{fig:glacier_masks}
\end{figure}

\begin{figure}
    \centering
    \includegraphics[width=0.5\linewidth]{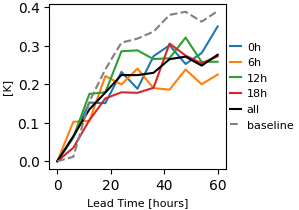}
    \caption{Difference between SFXG0 and SFX 2-metre temperature as function of forecast lead time. Each line represents forecasts initialized at different hours, black line represents all initialization times and dashed line shows the difference between BASELINE and the SFX reference.}
    \label{fig:glacier_temp_diff_lt}
\end{figure}

To determine whether the improvements are due to the provided glacier fraction, other forcings or by chance, we perform a set of extra inference runs during the month of highest errors in the BASELINE decoder (August 2022). The experiment includes a reference run with SFX and a run using SFX but with Glacier set to zero everywhere which we call SFXG0. Map of the resulting 2-metre temperature difference between the two runs is shown in Fig.~\ref{fig:glacier_masks}b. An increase in 2-metre temperature over the month is seen in the areas where glaciers are removed. The spatial average for different forecast lead times shown in Fig.~\ref{fig:glacier_temp_diff_lt}, indicates that 2-metre temperature differences increases rapidly with increasing forecast lead times up to 24 hours, then slower towards the end of the forecast. There is a slight shift between the curves for different initialization times. The forecasts initialized in the morning (06 UTC) increases most during the first time step, while the evening run (18 UTC) increases later. This suggests that the effect of removing glacier is strongest during daytime. We remind the reader that the initial 2-metre temperature field in each forecast is taken from MEPS which has similar glacier fractions as the reference run. 

\section{Discussion}
\label{seq:discussion}
In this study we target limitations in the description of surface conditions in current kilometre-scale data-driven weather prediction models. By training a set of new decoder components with additional surface descriptors we assess whether the model can better represent local atmospheric conditions that occur over specific surface types, not captured by the baseline configuration decoder. 

While the overall impact of including additional surface descriptors is small, our results indicate consistent improvement of both 2-metre temperature and 10-metre wind speed over all surfaces. The largest impact is seen in the decoders including the surface descriptors inherited from the MEPS model (SFX and COMBINED). Particularly, we find the descriptors representing limited extent areas like town, coastlines and glacier to help the decoder learn the local conditions not captured by BASELINE. Mean errors of 10-metre wind speed reveal that BASELINE systematically predicts more wind over sea along the Norwegian coast. These areas contain oil platforms which reports wind speed assimilated by the target model (MEPS). During the period of the training data, poor vertical correction of these wind measurements (from sensor height to 10-metre) caused unwanted effects of the assimilation as described in \citet{koltzow_how_2024}. The difference between MEPS and BASELINE suggests that the data-driven model generalizes sufficiently to ignore these features which in this case is beneficial. The same mechanism is likely causing the model to miss special processes over other local conditions, like those over towns and glaciers. BASELINE also struggle to distinguish between larger scale surface conditions like fraction of forest which impacts the 10-metre wind speed significantly. Providing the decoders with surface descriptors inherited from the surface model of MEPS resolves many of these challenges. 

Topography descriptors are included to provide the model with neighbourhood context. However, it does not show a large impact on the forecasting scores, except for mountains. Here, both 2-metre temperature and 10-metre wind speed are better predicted by TOPO (1\% and 2.5\% reduction in MAE respectively, compared to BASELINE) which also outperforms SFX. These areas have generally larger errors compared to other areas for both variables. Investigation of relative MAE between BASELINE and TOPO (not shown) gives no clear indication of which of the topo descriptors contributes most to the slightly improved scores. This suggests that the effect of each topography descriptor are connected across several forcings. 

Investigating the model behaviour over glaciers gives valuable insight in how the Glacier descriptor impacts the 2-metre temperature forecast. The fact that the model responds in a physical consistent way strengthens the general confidence in the model as a tool for geophysical research. The shift in 2-metre temperature increase between initialization times suggests that SFX captures diurnal dependencies of the energy budget, e.g. described in \citet{fernandez-castillo_impact_2025}. Furthermore, being able to correct or update surface descriptors is crucial for operational forecasting systems. Many of the surface descriptors used in this study are evolving, like Glacier, Town and Forest due to climate change and anthropogenic activities. Since the data-driven models are expensive to train but cheap to run, being able to modify input data instead of retraining the model is very beneficial. This also highlights the benefit of using physically based surface descriptors compared to learnable features, which is the other main alternative to capturing local effects. While not investigated directly in this study, the results suggest that providing more accurate surface descriptor data, the decoder could potentially perform better than the target data if it is based on outdated or erroneous surface datasets. 

In this study we show the importance of static surface descriptors in the data-driven weather forecasting model. There are also a number of dynamic surface conditions that impact the land atmosphere interactions. \citet{cho_deep_2026} includes soil temperature and moisture to a data-driven weather prediction model with similar architecture used in this work and found improved predictions of heat waves. Over the Nordics and at high latitudes in general, seasonal snow cover and deciduous forest cause considerable differences in the land atmosphere interaction during the year. In the current baseline model, the dynamic forcings that can help the model to distinguish summer from winter are Insolation and cos/sin Julian day. In addition, several input variables like surface temperature is part of the input and could help the model infer the seasonal context. Surface temperature could indicate snow cover, but a land surface at freezing point could also occur for bare ground. These two different conditions would lead to significantly different evolution of the day time 2-metre temperatures \citep[e.g.][]{fernandez-castillo_impact_2025}. Including the dynamic surface conditions as prognostic model variables is thus a strong candidate for future work in our model framework. Furthermore, if soil and vegetation variables are included as prognostic variables, we suggest that static surface descriptors should also be included as the dynamics of soil variables are strongly dependent on the soil texture and vegetation type \citep{decharme_local_2011, decharme_impacts_2016}.

Data-driven models trained on reanalysis and operational numerical weather prediction archives have shown promising results, but they are limited by the underlying model physics and its simplifications and assumptions. A different branch of data-driven weather prediction models base their training solely on observational datasets \citep{allen_end--end_2025}. This approach allows models to learn processes that are approximated or not parametrized at all in the physical models. While this is an appealing advantage, it comes with some significant challenges. In a physically based model, the representativeness of a variable is well defined. For example, 2-metre temperature represents a grid cell with known properties like area extent and surface types. On the other side, a point measurement of 2-metre temperature does not represent a grid cell, but rather that specific point which can be influenced by very small scale conditions like nearby vegetation or buildings. Given the heterogeneity of the land surface, the observation of 2-metre temperature is not sufficient alone. In order to generalize information from these kinds of observations, it might be crucial to provide accurate information about what they represent. Fortunately, such data exist at very high resolution, for example from the sentinel 2 platform \citep[e.g.][]{malinowski_automated_2020, karra_global_2021}. 

By only training the decoder and keeping encoder and processor constant, we are able to evaluate the impact of surface descriptors at a relatively low cost. Since the decoder transforms the relatively coarse scale latent space into higher resolution, can assume that the high resolution surface descriptors are most important in the decoding step. However, its likely beneficial to also include the surface descriptors in the encoder step. For example, a town could impact development of smaller scale weather systems very different than a forest or lake with the same surface temperature. 

\section{Conclusion}
\label{seq:conclusion}
In this study we show that providing additional surface descriptors in a data-driven weather prediction model improves the prediction of 2-metre temperature and 10-metre wind speed compared to a reference model. Highest impact is found over town, coastline and glaciers when information about these fields are provided as input. Smaller improvements are seen over mountain regions when using topography neighbourhood indices. Overall best performance is found when combining surface cover information inherited from the target model surface scheme and topography descriptors derived from the elevation dataset. 

A case study where glaciers are removed, shows an increase in predicted 2-metre  temperature in those areas. This strengthens our confidence in the physical consistency of the model. The finding suggests that evolving surface conditions, like the shrinking glaciers, can be treated by updating input fields without the need to retrain the model. This is crucial for operational forecasting systems in light of the much higher cost of training compared to running data-driven models. 

Our study highlights the importance of including surface descriptors in data-driven models, specifically when targeting near-surface variables at high spatial resolution. The isolated training of the decoder component allowed low cost evaluation of the input datasets. As the results of this study demonstrate the importance of the surface descriptors, they should be included as input to the full model chain including encoder and processor in future model configurations. This approach not only enhances forecasting accuracy but also provides a flexible framework for addressing changing surface conditions, making it highly relevant for future applications in operational weather forecasting. Our findings suggests that including time varying surface variables, like snow cover, could improve forecasts performance even further. We recommend including these in addition to static surface descriptors in future development of data-driven weather forecasting models. 

\section*{Competing interests}
The authors declare no competing interests influencing the work presented this article. 

\section*{Acknowledgements}
We acknowledge the EuroHPC Joint Undertaking for awarding the project EHPC-REG-2024R02-079 access to the EuroHPC supercomputer LEONARDO, hosted by CINECA (Italy) and the LEONARDO consortium through an EuroHPC Regular Access call. We thank the contributors of Anemoi for developing and supporting the open source framework. 

\bibliographystyle{apalike} 
\bibliography{references}
\end{document}